\documentclass[fleqn,usenatbib]{mnras}

\usepackage{graphics,epsf}
\usepackage{amsmath}                
\usepackage{amsfonts}               
\usepackage{amssymb}                
\usepackage{epsfig}                 
\usepackage{graphicx}

\usepackage{xcolor}
\definecolor{redak}{rgb}{0.9,0.15,0.05}

\def \kms{~\rm{km~s^{-1}}}

\def \cm{~\rm{cm}}
\def \s{~\rm{s}}
\def \km{~\rm{km}}

\def \K{~\rm{K}}
\def \g{~\rm{g}}

\def \AU{~\rm{AU}}

\def \yr{~\rm{yr}}

\def \mum{~\rm{\mu m}}

\def \rmModot{~\rm{M_\odot}}

\def \rmLodot{~\rm{L_\odot}}

\title[Type II ILOT]{Type II intermediate-luminosity optical transients (ILOTs)}

\author[A. Kashi and N. Soker]{
Amit Kashi$^{1}$\thanks{E-mail: \href{mailto:kashi@ariel.ac.il}{kashi@ariel.ac.il}}
and
Noam Soker$^{2}$\thanks{E-mail: \href{mailto:soker@physics.technion.ac.il}{soker@physics.technion.ac.il}}
\\
$^{1}$Physics Department, Ariel University, Ariel, POB 3, 40700, Israel \\
$^{2}$Department of Physics, Technion, Haifa 3200003, Israel\\
\\
}

\date{Accepted XXX. Received YYY; in original form ZZZ}

\pubyear{2016}

\begin{document}
\label{firstpage}
\pagerange{\pageref{firstpage}--\pageref{lastpage}}
\maketitle

\begin{abstract}
We propose that in a small fraction of intermediate luminosity optical transients (ILOTs) powered by a strongly interacting binary system, the ejected mass in the equatorial plane can block the central source from our line of sight. We can therefore observe only radiation that is reprocessed by polar outflow, much as in type~II active galactic nuclei (AGN).
An ejection of $M_{\rm ej,e}=10^{-4} \rmModot ~ (1 \rmModot)$ at 30 degrees from the equatorial plane and at a velocity of $v_{\rm e} = 100 \km \s^{-1}$ will block the central source in the NIR for about 5 years (500 years). During that period of time the object might disappear in the visible band, and be detected only in the IR band due to polar dust.
We raise the possibility that the recently observed disappearance of a red giant in the visible, designated N6946-BH1, is a type~II ILOT rather than a failed supernova. For this case we estimate that the ejected mass in the polar direction was $M_{\rm ej,p}\approx 10^{-3} \rmModot$. Our scenario predicts that this event should reinstate its visible emission in several decades.
\end{abstract}


\section{INTRODUCTION}
\label{sec:intro}

Intermediate-luminosity-optical-transients (ILOTs) are a heterogeneous group of stellar outbursts with a general luminosity between those of classical novae and supernovae (SNe; e.g., \citealt{Kasliwal2013}).
They have attracted much attention in recent years (e.g., \citealt{Ofeketal2016, Blagorodnovaetal2016, Tartagliaetal2016, Villaretal2016,  Smithetal2016a, Goranskijetal2016a}; limited to papers from the last year).

The nature of ILOTs is still not fully understood, and there is even disagreement on how to term the different subgroups. We use the term ILOTs \citep{Berger2009} to describe the group, and include under ILOTs, among other transients, giant eruptions of luminous blue variables (LBV), such as the Great Eruption of Eta Carinae \citep{HumphreysDavidson1994, DavidsonHumphreys2012}, merger-bursts, such as V838~Mon \citep{SokerTylenda2006} and OGLE-2002-BLG-360 \citep{Tylendaetal2013}, and pre-explosion outbursts that occur before core collapse supernova (CCSN) explosions \citep{Ofeketal2008}.
We note other names that are being used, Intermediate-Luminous Red Transients (ILRT; or Intermediate-Luminous Transients, e.g., \citealt{Humphreysetal2011}), Luminous Red Novae (or Red Novae, e.g., \citealt{Kasliwaletal2011}), and Red Transients (\citealt{Bond2011}). In \cite{KashiSoker2016a} we discussed the different names and subgroups.
The Energy-Time Diagram\footnote{See the ILOT-Club website with the ETD \newline \url{http://phsites.technion.ac.il/soker/ilot-club/}} (ETD) is a useful tool to characterize ILOTs , where the ILOT total energy is plotted versus their eruption duration.

The models and scenarios for the different subclasses of ILOTs include single star models
(e.g., \citealt{Thompsonetal2009, Kochanek2011, Ofeketal2013}),
and rich varieties of stellar binary interaction processes (\citealt{SokerTylenda2003, Kashietal2010, KashiSoker2010b, Tylendaetal2011, SokerKashi2011, SokerKashi2012, Ivanovaetal2013, SokerKashi2013, Tylendaetal2013, McleySoker2014, Nandezetal2014, Kaminskietal2015a, Kaminskietal2015b, Soker2015,  IvanovaNandez2016, Goranskijetal2016, Soker2016, MacLeodetal2016, Blagorodnovaetal2016, Pejchaetal2016a, Pejchaetal2016b, Smithetal2016b, Zhuetal2016}), and even interaction with planets \citep{RetterMarom2003, Retteretal2006, Bearetal2011, Metzgeretal2012}.

In many of the binary ILOTs the secondary star is at least partially responsible for the ejection of a large fraction of the envelope of the giant star in specific directions, in particular at low latitudes, namely, close to the equatorial plane. This has been shown in three-dimensional hydrodynamical simulations of the onset of the common envelope evolution (CEE), where an outflowing torus (ring) or an accretion disc might form (e.g., \citealt{LivioSoker1988, RasioLivio1996, SandquistTaam1998, Lombardi2006, RickerTaam2012, Passyetal2012, Nandezetal2014, Staffetal2016, Ohlmannetal2016, Iaconietal2016}).

In this paper we study cases where the column density of the outflow toward our line of sight, such as in the equatorial plane, is sufficiently large to block a large fraction of the radiation of the central binary system. This will result in a declining luminosity that might last tens of years. When the optically thick outflow is near the equatorial plane, radiation from the central source might be processed by gas and dust residing in the polar directions and emitted toward our line of sight, much is in type~II active galactic nuclei (AGN) where the broad line region is obscured.
In an analogy to type~II AGN we shall refer to these ILOTs as \textit{type~II ILOTS}. We note that the same ILOTs, observed from a higher latitude above the obscuring torus, would look different and would not be classified as type~II by the observer.

The idea that an equatorial outflow/disc can obscure the outburst of an ILOT was discussed before. 
\cite{Kaminskietal2010} suggested that V4332 Sag is a supergiant surrounded by a circumstellar disc, which is seen almost edge-on so that the central star is obscured from the observer. According to their model the observed light comes from scattering the supergiant spectrum on dust grains at the outer edge of the disc and possibly on a polar outflow (\citealt{Kaminskietal2010}; \citealt{KaminskiTylenda2013}).
\cite{Kaminskietal2015b} suggested a similar model for V1309 Sco.
We here extend and further explore this scenario, which we term a type II ILOT. 

In section \ref{sec:typeII} we describe a model for type~II ILOTs, and in section \ref{sec:LC} we present the expected light curve of Type II ILOT.
In section \ref{sec:N6946} we examine the possibility that the event N6946-BH1 was a Type~II ILOT rather than a failed CCSN. In the N6946-BH1 event there was an outburst, followed by a significant still-lasting decline in luminosity \citep{Adamsetal2016}.
We summarize our study in section \ref{sec:summary}.

\section{THE PROPOSED TYPE II ILOT}
\label{sec:typeII}

\begin{figure*}
\includegraphics[trim= 2.0cm 1.5cm 0.0cm 0.0cm,clip=true,width=0.98\textwidth]{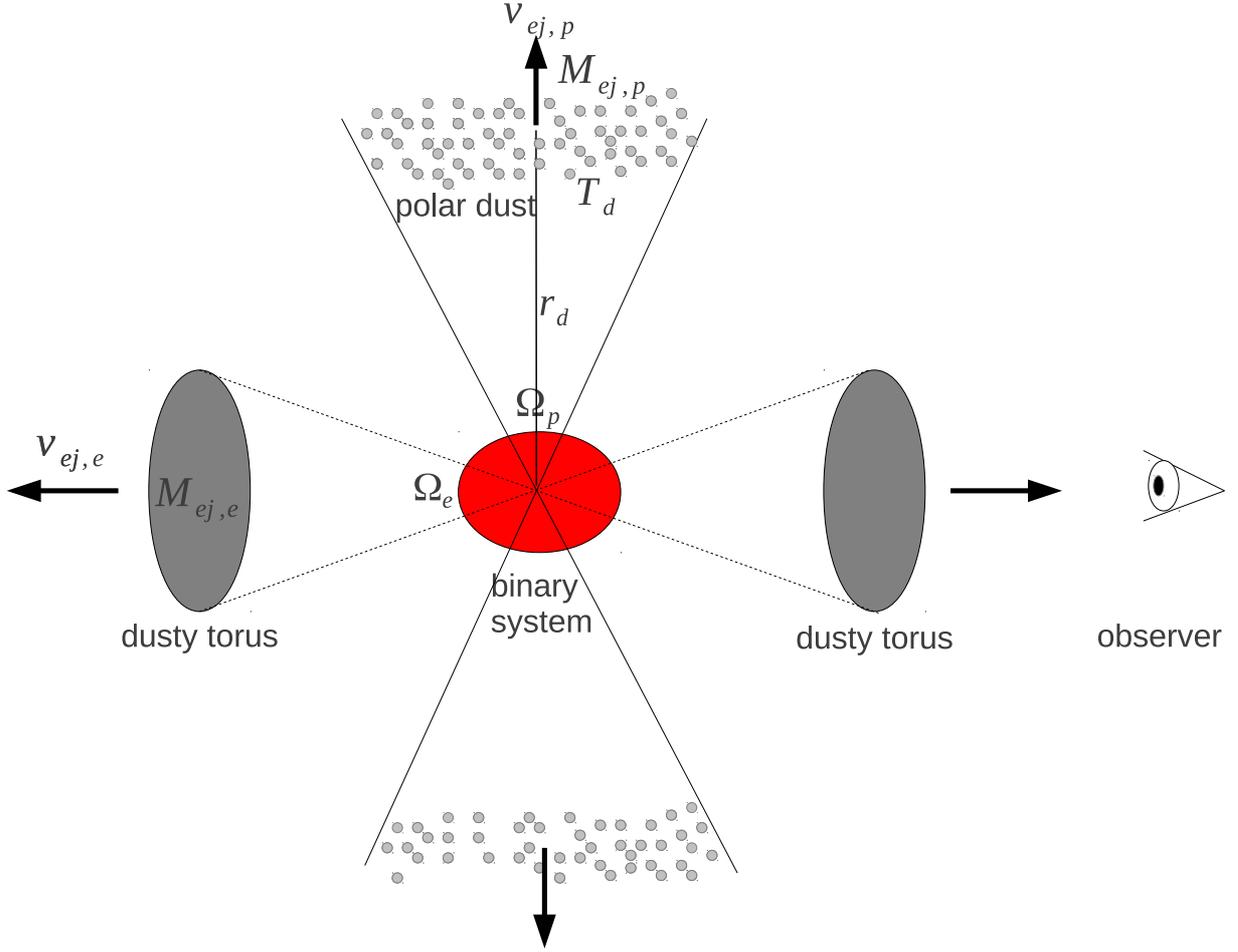}
\caption{A cartoon of the gas and dust outflow morphology resulting from a strong binary interaction. When the optical depth of the equatorial ejecta is very large and we happen to be close to that plane, the central light source is obscured, but not light reflected or emitted by the polar ejecta. This leads to a type~II ILOT.
}
\label{fig:cartoon}
\end{figure*}

The scenario we discuss is based on a strong binary interaction. Such an interaction can be a periastron passage similar to the Great Eruption of $\eta$ Carinae (\citealt{KashiSoker2010a}), or a terminal merger.
The interaction leads to an axisymmetrical mass ejection with a large departure from spherical symmetry, much like the morphologies of many planetary nebulae (e.g. \citealt{Balick1987, CorradiSchwarz1995, Manchadoetal1996, Sahaietal2011, Parkeretal2016}).
The properties of the gas and dust ejected in the equatorial plane will differ from the properties of the polar ejecta.
For that reason we will scale differently the expressions for the properties of the equatorial and polar ejecta.
In Fig. \ref{fig:cartoon} we present schematically the ejecta resulting from this strong binary interaction.
  
Consider then a strongly interacting binary system that ejects mass $M_{\rm ej,e}$ into a solid angle $\Omega_e = 4 \pi \delta_e$ around the binary equatorial plane with an outward terminal velocity of $v_{\rm e}$, as depicted in Fig. \ref{fig:cartoon}.
Similarly, it ejects mass $M_{\rm p}$ into a solid angle $\Omega_{\rm p} = 4 \pi \delta_{\rm p}$ around the polar directions, and with a terminal velocity of $v_{\rm p}$. We consider the interacting binary system and the outflow a long time after the main ejection even. Optically thin wind and jets can continue to be blown at later times.

The column density of the equatorial ejecta is give by the following expression, scaled to typical values of the proposed Type II ILOT scenario.
\begin{equation}
\begin{split}
\Sigma =  \frac {M_{\rm ej,e}}{4 \pi \delta_{\rm e} r^2}  & = 32
\left( \frac {M_{\rm ej,e}}{1 \rmModot} \right)
\left( \frac {v_{\rm e}}{100 \km \s^{-1}} \right)^{-2}    \\
& \times
\left( \frac {t}{10 \yr} \right)^{-2}
\left( \frac {\delta_{\rm e}}{0.5} \right)^{-1}    \g \cm^{-2}
\end{split}
\label{eq:Sigma}
\end{equation}
where $t$ is the time since the outburst.
For a solar composition this column density corresponds to an hydrogen number column density of $N_{\rm H}=1.3 \times 10^{25} \cm^{-2}$.

In a relatively short time the outflowing gas cools, and at these high densities dust rapidly forms.
The opacity in the optical bands becomes very high $\kappa \approx 1-10 \cm^2 \g^{-1}$.
The ILOT of a red giant (ILRT) might involve an ejected mass of
$M_{\rm ej} \approx 0.1-1 \rmModot$, e.g., NGC~300OT (\citealt{Berger2009}; \citealt{SokerKashi2012}).
The central source might stay obscured for up to tens of years in the optical and UV.

The IR suffers much less absorption. In addition, the dust that outflows along the polar directions is heated because it is exposed to the central source.
This dust can be a relatively strong source of IR radiation.
We first show that for some range of ILOT parameters even the IR might be obscured by the equatorial outflow.
The wavelength-dependent extinction in the near IR (NIR) and for wavelengths of $\lambda \ga 0.4 \mum$ is approximated by \citep{Mathis1990}
\begin{equation}
\frac{A(\lambda)}{A(J)} \approx \left(\frac{\lambda}{1.25 \mum}\right)^{-1.7},
\label{eq:alabdaoveraj}
\end{equation}
where $A(J)$ is the extinction at $\lambda=1.25 \mum$.
The column density relates to this ratio with the formula \citep{Mathis1990}
\begin{equation}
N_H = 8.7 \times 10^{21} A(J) \rm{cm^{-2}}
\label{eq:alabdaovernh}
\end{equation}
which was obtained for an optical total-to-selective extinction ratio $R_V=(A(B)-A(V))/A(V) = 3.1$,
for LMC carbon abundance.
 
Using the relation between the extinction and optical depth $A(\lambda)=1.086\tau(\lambda)$, and equations \ref{eq:Sigma}, \ref{eq:alabdaoveraj} and \ref{eq:alabdaovernh}, we derive 
the optical depth as function of time
\begin{equation}
\begin{split}
 \tau(\lambda) & 
 \approx 1800 \left(\frac{\lambda}{1.25 \mum}\right)^{-1.7} \left( \frac {t}{10 \yr} \right)^{-2} 
 \left( \frac {M_{\rm ej,e}}{1 \rmModot} \right) \\ &
 \times \left( \frac {v_{\rm e} \vphantom{\delta}}{100 \km \s^{-1}} \right)^{-2} \left( \frac {\delta_{\rm e}}{0.5} \right)^{-1} \end{split}
\label{eq:taulambda} 
\end{equation}
This means that the dusty wind becomes optically thin after
\begin{equation}
\begin{split}
t_{\rm thin,e} (\lambda) & \approx 425 \left(\frac{\tau}{1}\right)^{-1/2} \left(\frac{\lambda}{1.25 \mum}\right)^{-0.85}
\left( \frac {M_{\rm ej,e}}{1 \rmModot} \right)^{1/2} \\
& \times
\left( \frac {v_{\rm e} \vphantom{\delta}}{100 \km \s^{-1}} \right)^{-1} \left( \frac {\delta_{\rm e}}{0.5} \right)^{-1/2} \yr.
\end{split}
\label{eq:tlambda}
\end{equation}
For a much lower ejected mass of $M_{\rm ej, e} = 0.1 \rmModot$, for example, the IR luminosity at $5 \mum$ might stay completely obscured for an equatorial observer for decades.

The morphology of the ejected mass is important for the model.
The obscuration by a torus, or a disk, is not at all like that of a spherical shell (as long as the photons scattered and emitted by the equatorial outflow can escape from high-latitude directions).
In the case of a spherical shell the photons will eventually diffuse out.
So the observer will be able to see the radiation of the central source, even if at longer wavelengths, e.g., the far IR.
In the case of a torus, the majority of photons will eventually escape at high-latitude direction (polar directions).
An observer near the equatorial plane will receive a small fraction of the radiation of the equatorial matter, only the portion of the radiation that is reprocessed by the polar outflow. 

The polar outflow is expected to have different properties, in particular a higher velocity, as is the case for example in the bipolar nebula of $\eta$ Carinae, the Homunculus, \citep{Smithetal2003}.
We shall calibrate the polar outflow velocity with $v_{\rm p} = 300 \km \s^{-1}$ and its mass with ${M_{\rm p}} = 0.001 \rmModot$.
Namely, we assume that the majority of the mass was ejected in equatorial directions and only a small fraction of the total ejected mass flows along the polar directions. As well, the polar outflow might have existed before the outburst. The flowing dust, especially the dust closer to the star, absorbs some of the stellar radiation and re-emits it in the IR.
For the typical parameters of the proposed scenario, the polar outflow becomes optically thin in the visible at a time of 
\begin{equation}
\begin{split}
t_{\rm thin,p}(\lambda) & \approx 15.5 \left(\frac{\tau}{1}\right)^{-1/2} \left(\frac{\lambda}{0.5 \mum}\right)^{-0.85}
\left( \frac {M_{\rm ej,p}}{0.001 \rmModot} \right)^{1/2} \\
& \times
\left( \frac {v_{\rm p} \vphantom{\delta}}{300 \km \s^{-1}} \right)^{-1} \left( \frac {\delta_{\rm p}}{0.2} \right)^{-1/2} \yr.
\end{split}
\label{eq:thinp}
\end{equation}
The luminosity of the polar material is 
\begin{equation}
L_{\rm p} \ga \delta_{\rm p} L_{\rm cen} \left(1- e^{-\tau_{\rm p}}\right),
\label{eq:lp}
\end{equation}
where $\tau_{\rm p}$ is the optical depth of the polar ejecta in the radiation band of the central source, and $L_{\rm cen}$ is the luminosity of the central source. The reason for the $\ga$ sign instead of an equal sign is that the polar ejecta reprocesses some of the radiation emitted by the equatorial ejecta in addition to that of the central source.
   
To model the dust we take the a modified blackbody function with a blue excess
\begin{equation}
\tilde{B}(T)= Q B(T) = \left(\frac{\lambda}{\lambda_0}\right)^{-m} B(T),
\label{tildeB}
\end{equation}
where $B(T)$ is the usual Planck function, and $m=1$ for amorphous
Carbon grains \citep{Hildebrand1983}.
We shall assume spherical dust grains with
radius $a = 0.25 \mum$ and density $\rho_d\sim 1.9 \g\cm^{-3}$
(\citealt{Kruegel2003, Whittet2003}). The ratio $Q/a$ is obtained
from figure 8.1 of Krugel (2003), by approximating the wavelength
range $1 \mum \leq \lambda \leq 100 \mum$
\begin{equation}
\frac{Q}{a} \simeq \frac{7.93}{\lambda}.
\label{Q1}
\end{equation}
For an assumed grain radius of $a \sim 0.25 \mum$
we find $\lambda_0 \simeq 2 \mum$.
The thermal equilibrium of the dust at distance $r_d$ from the star satisfies (e.g, \citealt{Kruegel2003})
\begin{equation}
\sigma \frac{L_{\rm cen}}{4 \pi r_d^2}=4 \pi a^2 \cdot \pi c
\int_0^\infty \tilde{B}(T_d) \lambda^{-2} \,d\lambda,
\label{td1}
\end{equation}
where $T_d$ is the dust temperature and $\sigma$ is the Stefan-Boltzmann constant.
After integrating over the modified blackbody, we get
\begin{equation}
\begin{split}
T_d &= 385
\left(\frac{L_{\rm cen}}{3\times10^5 \rmLodot}\right)^{1/5}
\left( \frac {v_{\rm p} \vphantom{\delta}}{300 \km \s^{-1}} \right)^{-2/5} \\
& \times \left( \frac {t}{10 \yr} \right)^{-2/5} \K.
\end{split}
\label{td2}
\end{equation}
The above calculation, though done for specific dust properties is generally valid for any kind of dust the results would need rescaling). It shows that the radiation reprocessed by the polar outflow will be strong in the IR, i.e. at few$\times \mum$ to over $10 \mum$. 

Silicate dust have higher albedo than carbon dust (e.g., \citealt{DraineLee1984}). This implies that silicate dust scatter more light in the original wavelengths. Carbon dust particles reprocess more of the radiation, namely, carbon dust has a higher ratio of IR to visible than silicate dust. The steep decline in the visible light from N6946-BH1 (see section \ref{sec:N6946}), might imply that for our proposed scenario to work for N6946-BH1, the dust should be carbon-rich.

\section{THE LIGHT CURVES}
\label{sec:LC}

The ingredients necessary for an ILOT event to be of Type~II are as follows. 
(1) An ejecta along the line of sight that substantially obscures the central source.
Above we took the ejecta to be in and near the equatorial plane, as expected in a strong binary interaction.
(2) A large solid angle through which the photons can freely leave the system.
(3) The presence of gas and/or dust that has a clear view of the central source, and that is not obscured to the observer.
We take this material to be in the polar directions. In such a flow the overall expected light curve is determined by the different radiation sources. 
These conditions imply that type~II ILOTs are rare events. 

\begin{figure*}
\includegraphics[trim= 0.0cm 0.0cm 0.0cm 0.0cm,clip=true,width=0.98\textwidth]{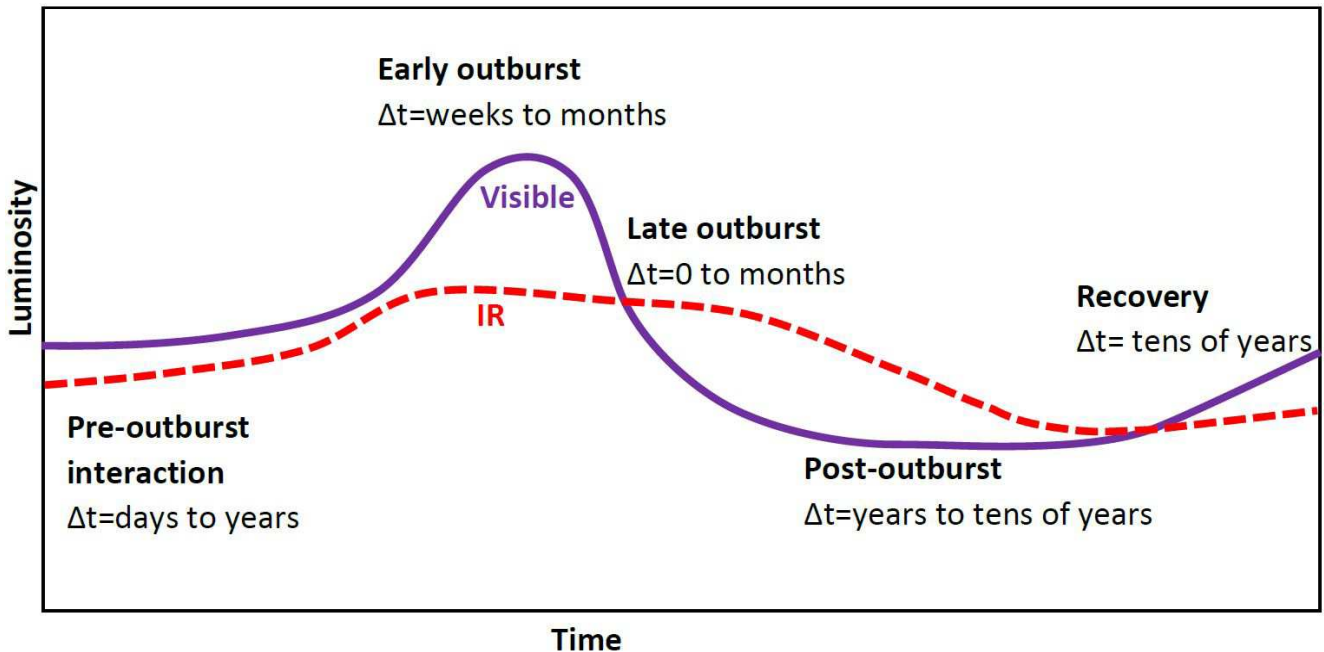}
\caption{A schematic drawing of the expected light curve of a type~II ILOT in the visible (purple) and in the IR (red). We distinguish between 5 phases (see text).
}
\label{fig:LC}
\end{figure*}

We can crudely divide the lightcurve to five phases. These are presented schematically in Fig. \ref{fig:LC}.
 
(1) \emph{Pre-outburst interaction.} As the interaction between the two stars starts, e.g., by the secondary accreting mass from the wind of the primary star, and/or by a disturbance to the primary star caused by the secondary star, the luminosity increases.
Gas is being ejected in the equatorial direction, and when it cools (both adiabatically and radiatively) dust starts to form.
The equatorial dust does not fully obscure yet the central source. The luminosity in the visible might increase because of the interacting central binary system.
The dust that already exists in the system and the newly formed dust reprocess some of this radiation to the IR band, and the IR luminosity increases. If the relative dust obscuration increases more than the increase in luminosity, the luminosity in the visible (and UV) will decrease. Over all, the ratio of IR to visible luminosity might increase in the pre-outburst interacting phase. This phase might be very short if the strong interaction occurs in an abrupt manner. 

(2) \emph{Early outburst.} In the proposed scenario a large amount of mass is ejected near the equatorial plane during the outburst. The process might even be the onset of a common envelope phase (e.g., \citealt{Ivanovaetal2013}). The gas cools and due to the high density, dust rapidly forms. Before the formation of dust the cooling gas emits radiation that comes from its initial thermal energy and from recombination. This radiation increases the total luminosity. Not all of the thermal energy and recombination energy is radiated away. Some fraction of the thermal and recombination energy might end in kinetic energy of the ejecta as it is absorbed in the gas, much as has been proposed in the process of the ejection of a common envelope at early phases (e.g., \citealt{Ivanovaetal2013}). In any case, after the gas recombines and before dust forms the optical depth is relatively low, and a large fraction of the recombination energy can be radiated away. Over all, the luminosity increases, both in the visible and in the IR bands.  This phase will last for a typical time about equal to the dynamical time of the strongly interacting binary system.

Consider for example a recombining solar-composition gas of mass $1 \rmModot$ that is ejected from a giant of several solar masses and of a radius of about $1 \AU$. The recombination phase lasts for about a month with a luminosity of $\approx 10^6 \rmLodot$, much as for common envelope ejection (e.g., \citealt{Ivanovaetal2013}).

Few words are in place here about the recombination energy. Firstly, the radiated energy is not the original UV photons from the recombination process, but it is rather the radiation that has been processed below the photosphere. Secondly, although the outer parts of the envelope of red giants are only partially ionized, in the proposed scenario a large amount of mass is ejected, such that most of it originates in the ionized layers of the envelope. Thirdly, the strong binary interaction heats up the gas and can further ionizes it. The gas can be heated by the friction as it flows around the stellar companion before it is ejected from near the second Lagrangian point. Else, parcels of gas ejected at different velocities might collide and heat up. 

(3) \emph{Late outburst.}  Interaction continues, but now the dust in the equatorial plane already obscures the central source. Visible luminosity substantially decreases. If polar outflow exists, the dust there sees the central source and emits in the IR band. The observed IR emission does not decrease much, and might even increase if the central binary system becomes more luminous.  
If the strong binary interaction ends before the equatorial ejecta obscures the central source, this phase does not exist. So it might last up to the typical dynamical time of the system.  
  
If the polar outflow is optically thin to visible radiation, then a large fraction of the observed light will be scattered visible light (depending on the dust albedo). If the polar outflow is optically thick, then most of the scattered visible light will be reprocessed and most of the observed radiation will be in the IR, according to the dust temperature. The later case better fits the observed properties of N6946-BH1. 

(4) \emph{Post outburst.} The interaction ceases and luminosity decreases. The equatorial ejecta continues to obscure the central binary source (or merger product). The dominate source of radiation is the dust along the polar directions. The system is observed to be much brighter in the IR than in the visible.

(5) \emph{Recovery.} At a time given approximately by equation (\ref{eq:tlambda}), the optical depth decreases and we start to observe the central source directly, first in the IR and then in the visible.

This description might change if the ejecta collides with slower circumstellar matter that was blown before the outburst. Such a collision heats the gas and increases the luminosity, both in the visible and IR.
    
\section{A TYPE II ILOT MODEL FOR N6946-BH1}
\label{sec:N6946}

The star N6946-BH1 erupted in 2009 \citep{Gerkeetal2015} in what was suggested to be a failed supernova event \citep{Adamsetal2016}.
The progenitor according to \cite{Adamsetal2016} is a $\approx 25 \rmModot$ red supergiant star with a radius of $\approx 2 \AU$. The Keplerian orbital period on the surface of such a star is about half a year.
About three years before the outburst the visible luminosity started to decrease slowly, while the NIR luminosity (observed in $3.6\mum$ and $4.5 \mum$) increases slowly
(see figure 2 of \citealt{Adamsetal2016}).
The star experienced then an outburst that lasted for about one year in the visible with a total luminosity of $> 10^6 \rmLodot$. At the end of the outburst and within several months the visible luminosity dropped by a factor of several hundred, to the extent of disappearance of the star.

The NIR has been decaying on a timescale of several years, and only by a factor of $<10$.
The dust responsible for the NIR emission has cooled over these years.
\cite{Adamsetal2016} attribute the residual NIR emission to accretion of mass onto the central black hole (BH) remnant. 
 
\cite{Adamsetal2016} favorite model for the outburst was a failed supernova event quite similar to the model of \cite{LovegroveWoosley2013}, who, based on an earlier ideas of \cite{Nadezhin1980}, simulated the hydrodynamic response of the star to a sudden loss of mass via neutrinos as the core of a red supergiant forms a protoneutron star.
\cite{Adamsetal2016} also examined several alternative models for the outburst, and concluded that the transient event is unlikely to be a SN impostor or a stellar merger.
In their study of alternative scenarios \cite{Adamsetal2016} assumed that the event was spherically symmetric.
This assumption was very limiting. However, this is unlikely to be the case in strongly interacting binary systems We here raise the possibility that N6946-BH1 was a type~II ILOT event rather than a failed supernova.
 
\cite{Adamsetal2016} concluded that if the star survived it cannot be hidden by a dusty wind because the hot dust that dominates
the obscuration re-radiates the stellar emission in the near to mid-IR.
They noted that in order to hide the luminosity of the progenitor, most of the emission must be radiated by cooler dust at wavelengths
redward of $4.5 \mum$, the longest wavelength measured.
This constrain however, is, again, only relevant for spherically symmetric ejecta, and not for directional outburst as we suggest here.

\cite{Adamsetal2016} constrained the velocity of the outer ejecta to $170 <v< 560 \kms$. 
The escape velocity of the progenitor with the mass and radius mentioned above is $v_{\rm esc} \simeq 150 \kms$.
Typically mass is ejected with a velocity of 1--3 times the escape velocity.
We therefore calibrate our scenario with $v_{\rm p}=200\kms$.
For $\delta_{\rm p}=0.2$ (a half-opening angle of 37 degrees), we get according to equation (\ref{eq:thinp})   
\begin{equation}
\begin{split}
\tau & \approx 0.5\left(\frac{t_{\rm p}}{10 \yr}\right)^{-2} \left(\frac{\lambda}{0.5 \mum}\right)^{-1.7}
\left( \frac {M_{\rm ej,p}}{10^{-4} \rmModot} \right) \\
& \times
\left( \frac {v_{\rm p} \vphantom{\delta}}{200 \km \s^{-1}} \right)^{-2} \left( \frac {\delta_{\rm p}}{0.2} \right)^{-1} .  
\end{split}
\label{eq:tau2}
\end{equation}

{}From figure 2 of \cite{Adamsetal2016} we see that $\approx 7 \yr$ after the event the $4.5\mum$ luminosity had decreased to $\approx 10^4 \rmLodot$. We can fit this with the following demonstrative values. 
The central source returned to a little below its pre-outburst luminosity, $L\approx 10^5 L_\odot$, the polar outflow has ${\delta_{\rm p}}={0.2}$, a velocity of $v_{\rm p} ={200 \km \s^{-1}}$, and the polar ejected mass is ${M_{\rm ej,p}} \approx 0.6 \times 10^{-4} \rmModot$. From equation (\ref{eq:lp}) we find the polar luminosity to be $\approx 10^4 \rmLodot$.   

We note that the  IR emitted by the polar outflow can decrease both as a result of decreasing optical depth of the polar ejecta (hence it absorbs less from the central source radiation), and as a result of a decrease in the luminosity of the central star. A continuous polar outflow will further complicate this calculation. Therefore, we cannot determine the relative contribution of each of the different factors. 
A decrease in the optical depth of the polar outflow will result in an increasing portion of the directly scattered light from the central source. Namely, the fraction of observed light at shorter wavelength, in particular in the visible, will increase.

If our model is correct, observations of N6946-BH1 in longer wavelengths may reveal that the star has survived and is hidden by cool dust.

We note that some (but not all) of the properties of type II ILOTs can be accounted for by an expanding spherical shell. 
In principle, a hot dust that is formed in a spherical outflow can obscure the ILOT in the visible, but the object remains bright in the near-IR. With time dust cools off and the near-IR flux decreases. A decrease in the near-IR was observed in the late evolution of V1309 Sco \citep{TylendaKaminski2016} and of OGLE-2002-BLG-360 \citep{Tylendaetal2013}, that were otherwise quite similar to N6946-BH1, in the rapid drop in the optical brightness that was accompanied by a significant rise in the IR. Although in V1309~Sco the flow was probably asymmetric, with a dense equatorial outflow, the observed IR evolution can be fully accounted for by a spherical expanding and cooling mass and dust.  As stated by \cite{Adamsetal2016}, such a model would result in a too strong IR emission in the case of N6946-BH1, and therefore cannot be applied to this specific ILOT.
 
\section{SUMMARY}
\label{sec:summary}

We examine a model for a new type of binary-powered ILOTs, and term them type~II ILOTs.
The event may be similar to regular ILOTs (type~I), but different in the orientation of the observer, whose line of sight crosses a thick dust shell or torus that obscures a direct view of the binary system or the merger product.
The obscuring matter would be in most cases equatorial, and will contain most of the ejected mass. 
Until this dust disperses the  binary system will be blocked to the observer in the visible and IR bands.
The event is accompanied by some polar mass ejection that also forms dust. The polar dust and gas reprocess the radiation from the central source, hence allowing the observation of the type~II ILOT, which becomes much fainter. A resembling scenario was discussed before by \cite{Kaminskietal2010} and \cite{KaminskiTylenda2013}. 

\cite{Adamsetal2016} observed a huge decrease in the visible luminosity (to the extent of disappearance) of the red supergiant N6946-BH1 in 2009. They gathered observations of the object in the visible and NIR for 7 years.
\cite{Adamsetal2016} examined a number of scenarios for this event, including different variable stars, SN~impostor and a merger event.
Their preferred model is a failed core-collapse SN that created a BH but did not produce much luminosity, as predicted by \cite{LovegroveWoosley2013}.
Their hope was that the discovery of this object will solve the problem of the missing
high-mass SN progenitors.
The scenarios examined in \cite{Adamsetal2016} were all spherically symmetric. Therefore, they did not allow part of the radiation to be directed out of the observer's line-of-sight.

We examined the possibility that the eruption of N6946-BH1 was an intermediate luminosity optical transient event (ILOT). We suggest that it was a type-II ILOT, namely that it became obscured from our line of sight as its ejecta formed a dense dusty torus. Our scenario predicts that the star survived and that in a several decades the dusty torus will dissipate and the visible emission will increase again.

The main claim of our paper is that in rare cases ILOT events might be observed as disappearing star for years to tens of years.

\section*{Acknowledgements}
We thank Ari Laor for helpful discussions. 
We thank an anonymous referee for helpful comments.
NS acknowledges support by the Israel Science Foundation.

\label{lastpage}
\end{document}